\documentclass[reprint,amsmath,amssymb,aps,superscriptaddress]{revtex4-1}

\usepackage{graphicx}
\usepackage{color}

\begin{document}

\title{Tuning 2D hyperbolic plasmons in black phosphorus}

\author{Edo van Veen}
\affiliation{School of Physics and Technology, Wuhan University, Wuhan 430072, China}
\affiliation{Institute for Molecules and Materials, Radboud University, Heyendaalseweg 135, 6525AJ Nijmegen, The Netherlands}
\author{Andrei Nemilentsau}
\affiliation{Department of Electrical and Computer Engineering, University of Minnesota, Minneapolis, Minnesota 55455, United States}
\author{Anshuman Kumar}
\affiliation{Physics Department, Indian Institute of Technology Bombay, Mumbai 400076, India}
\author{Rafael Rold\'an}
\affiliation{Materials Science Factory. Instituto de Ciencia de Materiales de Madrid (ICMM), Consejo Superior de Investigaciones Cient\'{i}ficas (CSIC), Cantoblanco E28049
Madrid, Spain}
\author{Mikhail I. Katsnelson}
\affiliation{Institute for Molecules and Materials, Radboud University, Heyendaalseweg 135, 6525AJ Nijmegen, The Netherlands}
\author{Tony Low}
\email{tlow@umn.edu}
\affiliation{Department of Electrical and Computer Engineering, University of Minnesota, Minneapolis, Minnesota 55455, United States}
\author{Shengjun Yuan}
\email{s.yuan@whu.edu.cn}
\affiliation{School of Physics and Technology, Wuhan University, Wuhan 430072, China}
\affiliation{Institute for Molecules and Materials, Radboud University, Heyendaalseweg 135, 6525AJ Nijmegen, The Netherlands}

\date{\today}

\begin{abstract}
Black phosphorus presents a very anisotropic crystal structure, making it a potential candidate for hyperbolic plasmonics, characterized by a permittivity tensor where one of the principal components is metallic and the other dielectric. Here we demonstrate that atomically thin black phosphorus can be engineered to be a hyperbolic material operating in a broad range of the electromagnetic spectrum from the entire visible spectrum to ultraviolet. With the introduction of an optical gain, a new hyperbolic region emerges in the infrared. The character of this hyperbolic plasmon depends on the interplay between gain and loss along the two crystalline directions. 

\end{abstract}

\maketitle

\emph{Introduction.}
Semiconducting two dimensional (2D) crystals are excellent platforms for tuneable optoelectronics, thanks to their remarkable response to external electrical and mechanical stimuli \cite{Avouris_Book_2017,Roldan_CSR_2017}. In particular, atomically thin black phosphorus \cite{Li_NN_2014,liu2014phosphorene,Castellanos-Gomez_2DM_2014,Xia_NC_2014} (BP) has shown extraordinary tuneability of its optical and electronic properties by several methods \cite{Roldan_NP_2017}, like electrostatic gating \cite{lin2016multilayer,peng2017midinfrared,whitney2016field,Deng_NC_2017,Liu_NL_2017}, chemical functionalization \cite{Kim_S_2015}, quantum confinement (number of layers) \cite{Yang_LSA_2015}, external strain \cite{quereda2016strong} or high pressure \cite{Rodin_PRL_2014,Xiang_PRL_2015}. This allows the control of light-matter interaction in these materials, in particular the dispersion of collective polaritonic excitations \cite{Low_NM_2017}. 

Apart from being a highly tuneable optoelectronic crystal, the lattice structure of BP is very anisotropic \cite{Avouris_Book_2017,keyes1953electrical}. The in-plane anisotropy implies optical birefringence, of which the extreme limit would be hyperbolicity, where the  permittivity tensor has principal components of opposite sign \cite{Poddubny_NP_2013,yermakov2015hybrid,gomez2015hyperbolic,Nemilentsau_PRL_2016}. Recently, in-plane hyperbolicity was implemented experimentally in the GHz frequency range using a metallic metasurface \cite{yermakov2018experimental}. Moreover, in-plane hyperbolicity in natural van der Waals material $\alpha$-MoO$_3$ was reported and existence of hyperbolic surface polaritons was experimentally verified \cite{ma2018plane,zheng2018mid}. The strong anisotropy of BP suggests its potential as a natural hyperbolic material, offering new possibilities for actively manipulating polaritons in 2D, such as directional plasmons, light emitters, superlensing effects, \cite{Nemilentsau_PRL_2016,Correas-Serrano_JO_2016} etc. In this Letter, we discuss the possiblity of driving BP into the hyperbolic region via electrostatic tuning, strain or layers number. We demonstrate that atomically thin BP can be efficiently tuned to become  hyperbolic in a broad spectral range from the entire visible spectrum to the ultraviolet. In addition, the presence of a bandgap in excess of the optical phonon energy lends itself as a possible 2D semiconductor gain medium \cite{shang2017room,wu2015monolayer}. 
With the introduction of a population inversion, we show that optical gain results in a new hyperbolic region in the infrared. Finally, we study the behavior of plasmons in both of these hyperbolic regions.

\emph{Optical Conductivity and Band Model.}
We describe BP by means of a $p_z$-orbitals tight-binding model fitted to {\it ab initio} $GW$ methods \cite{rudenko2014quasiparticle, rudenko2015toward}.

\begin{equation}
    \mathcal{H}=\sum_{i\neq j}t_{ij}c_{i}^{\dagger}c_{j}+\sum_{i\neq j}t_{p,ij}c_{i}^{\dagger }c_{j}, 
\end{equation}
where $c_{i}^{\dagger}$ ($c_{i}$) creates (annihilates) an electron at site $i$, and ten intra-layer $t_{ij}$ and five inter-layer $t_{p,ij}$ hopping terms are considered in the model. The obtained band structure corresponds to an anisotropic direct band gap semiconductor, with the gap at the $\Gamma$ point of the Brillouin zone. 

The model can be straightforwardly extended to incorporate arbitrary electrostatic and strain fields. An electric field is applied by modifying the on-site potentials with $\epsilon_i = e \times \Delta U \times z_i$, where $e$ is the elementary charge, $\Delta U$ is the bias voltage (with units V/nm) and $z_i$ is the $z$-coordinate of site $i$.

On the other hand, the application of external strain leads to a variation of the interatomic bond lengths, which further modifies the hopping terms as \cite{Suzuura_PRB_2002}
\begin{equation}
    t_{ij}({\bf r}_{ij})=t_{ij}({\bf r}_{ij}^0)
    \left(1-\beta_{ij}
    \frac{|{\bf r}_{ij}-{\bf r}_{ij}^0|}{|{\bf r}_{ij}^0|}
    \right),
\end{equation}
where $|{\bf r}_{ij}^0|$ is the distance in the equilibrium positions
between two atoms $i$ and $j$,
$|{\bf r}_{ij}|$ the distance in the presence of strain,
and $\beta_{ij}=-d\ln t_{ij}(r)/d\ln(r)|_{r=|{\bf r}_{ij}^0|}$
is the dimensionless local electron-phonon coupling.
A microscopic estimation of 
$\beta_{ij}$
can be done based on the direct comparison
between  {\it ab initio} and tight-binding calculations. Here we use $\beta\approx 4.5$ because this value was proved to give a matching between {\it ab initio} and tight-binding calculations for the direct-to-indirect bandgap transition under uniaxial strain \cite{san2016inverse}. The mechanical properties of BP are highly anisotropic, with zigzag direction being about four times stiffer than armchair direction \cite{Wei_APL_2014,quereda2016strong}. Therefore we use uniaxial strain along the armchair direction for our calculations, accounted for by the strain tensor $\mathbf{\epsilon}_\mathrm{AC}=\epsilon_{yy}\mathrm{diag}(\nu^\mathrm{AC}_z, 1, -\nu^\mathrm{AC}_z)$, where the Poisson ratios $\nu$ are estimated to be $\nu^\mathrm{AC}_x\approx 0.2$, and $\nu^\mathrm{AC}_z\approx 0.2$ \cite{Wei_APL_2014}. We notice the importance to consider the out-of-plane Poisson ratio $\nu_z$ in our calculations, that accounts for the widening (flattening) of the lattice under compressive (tensile) strain.

The optical conductivity in the zigzag ($\sigma_{xx}$) and armchair ($\sigma_{yy}$) directions is given by the Kubo formula

\begin{eqnarray}
	\text{Re}(\sigma_{\alpha \alpha}(\omega)) = -\frac{g_S}{\Omega \omega} \int_{BZ}  \text{ Im} \left[ \sum_{i,j} |\langle \mathbf{k} i | J_{\mathbf{k} \alpha} | \mathbf{k} j \rangle| ^2 \right. \nonumber \\
	\left. \times \frac{f(E_{ \mathbf{k}i}- \mu) - f(E_{ \mathbf{k}j}- \mu)}{E_{ \mathbf{k}i} - E_{ \mathbf{k}j} + \omega + i \delta} \vphantom{\sum_{i,j}} \right] d^2 \mathbf{k} .
\end{eqnarray}
Here, $| \mathbf{k} i \rangle$ and $E_{ \mathbf{k}i}$ are the eigenstates and eigenenergies for momentum $\mathbf{k}$ and orbital $i$. $g_S=2$ is the spin degeneracy, $\Omega$ is the unit cell surface, and $\delta=5$ meV is a small damping parameter. $J_{\mathbf{k} \alpha}$ is the current operator in the $\alpha$-direction

\begin{equation}
	J_{\mathbf{k} \alpha} = - \frac{i e}{\hbar} \sum_{i, j} e^{i (\mathbf{r}_j - \mathbf{r}_i)_{\alpha} \cdot \mathbf{k}} t_{ij}  (\mathbf{r}_j - \mathbf{r}_i)_{\alpha} c_{\mathbf{k}i}^{\dagger} c_{\mathbf{k}j}^{\phantom{\dagger}}.
\end{equation}

Moreover, $f(E-\mu)$ is the Fermi-Dirac distribution with Fermi level $\mu$: 
\begin{equation}
    f(E-\mu) = \frac{1}{e^{(E-\mu)/kT} + 1},
\end{equation}
where we set $T=300$ K.

We can replace $f(E)$ with a quasi-equilibrium distribution $n_F(E, \Delta \mu)$ to introduce population inversion, which will produce optical gain \cite{low2018superluminal,page2015nonequilibrium}:

\begin{equation}
    \label{Eq:gain}
    n_F(E) = \theta(E) f(E + \frac{E_g}{2} + \Delta \mu) + \theta(-E) f(E-\frac{E_g}{2}-\Delta \mu),
\end{equation}
where $E_g$ is the band gap.

Using the Kramers-Kronig relations, we can also obtain the imaginary part

\begin{equation}
    \text{Im}(\sigma_{\alpha\alpha}(\omega)) = -\frac{2\omega}{\pi} \mathcal{P} \int_0^{\infty} \frac{\text{Re}(\sigma_{\alpha\alpha}(\omega'))}{\omega'^2 - \omega^2}d\omega'.
\end{equation}

\begin{figure}[t]
\includegraphics[width=\linewidth]{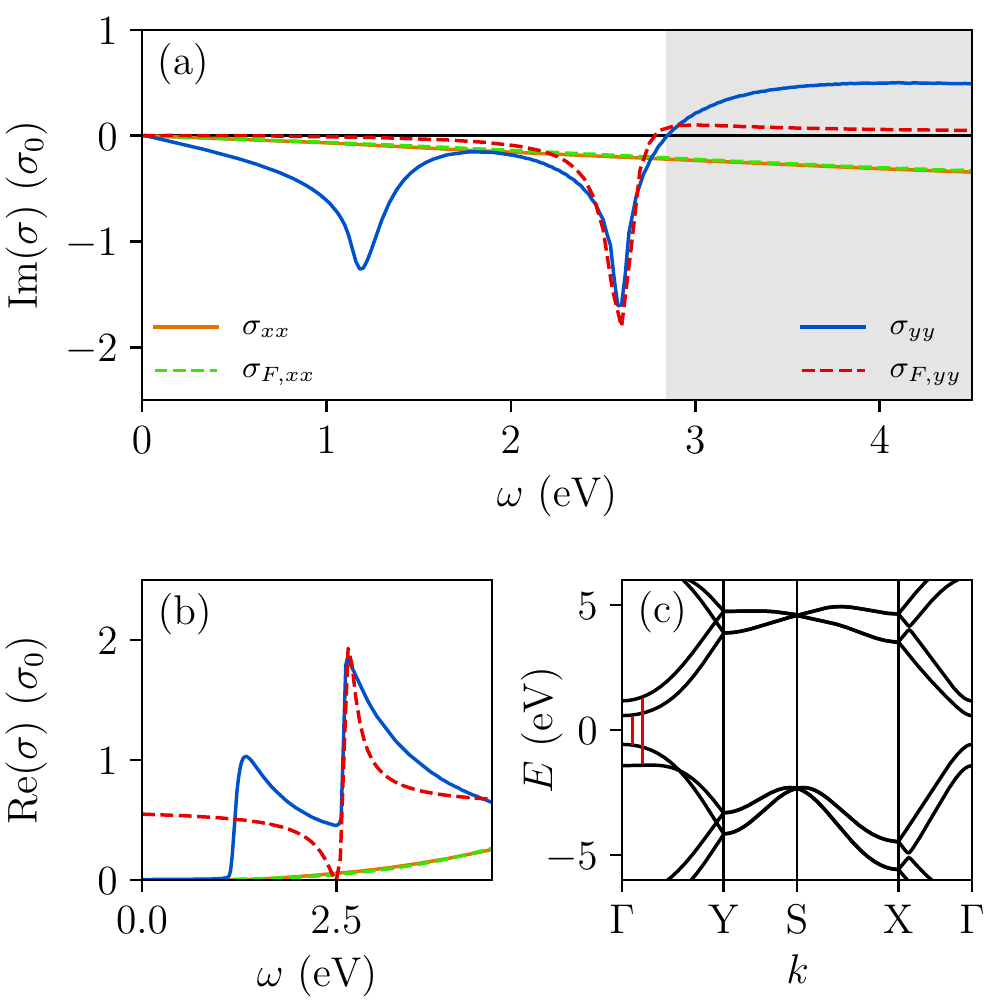}
\caption{(a) The imaginary part of the optical conductivity of bilayer black phosphorus in units of $\sigma_0 = \frac{\pi e^2}{2 h}$, showing a hyperbolic region (shaded) where $\text{Im}(\sigma_{xx}) \times \text{Im}(\sigma_{yy}) < 0$, starting at $\omega = 2.8$ eV. (b) The real part of the conductivity corresponding to (a). The dashed lines show a fit of the conductivity around the second peak, using the Fano model as described in the text, with resonance width $\Gamma_{res} = 0.1$ eV, Fano parameter $q_F=1.5 \text{\ eV}^{-1}$ and $n=3$. (c) The corresponding band structure, with optical excitations indicated in red, causing the two optical peaks at $\omega=1.2$ eV and $\omega=2.6$ eV.
}
\label{Fig:accond_pristine}
\end{figure}

\emph{Hyperbolic Regions.}
We first define the condition for hyperbolicity. We note that the real part of the dielectric permittivity is proportional to $\text{Im} (\sigma)$, a consequence of current continuity. Then, a hyperbolic region appears when
\begin{equation} \label{Eq:hyperbolic_condition}
    \text{Im} (\sigma_{xx}(\omega)) \times \text{Im} (\sigma_{yy} (\omega)) < 0.
\end{equation}
On the other hand, $\text{Re} (\sigma)$, is directly proportional to the optical absorption of the free-standing 2D layer.
For pristine bilayer BP, the optical conductivity components are plotted in Fig. \ref{Fig:accond_pristine}. The first thing we observe is that the peculiar puckered structure of BP leads to a strong linear dichroism, i.e., a large difference in optical conductivity for incident polarized light along armchair and zigzag directions \cite{Low_PRB_2014}. For a bilayer sample, its optical absorption revealed two sharp peaks along the armchair direction, due to the two interband excitations indicated in red in the band structure. These resonant-like features, for light polarized along armchair, has also been observed experimentally \cite{zhang2017infrared}. On the contrary, light polarized along zigzag shows a featureless monotonically increasing optical absorption instead.
Whereas $\text{Im}(\sigma_{xx})$ is negative throughout the spectrum, $\text{Im}(\sigma_{yy})$ goes from negative to positive around $\omega_h = 2.8$ eV, which results in a hyperbolic region starting at that frequency.

The sign change in $\text{Im}(\sigma_{yy})$ along the armchair direction is key to the appearance of the hyperbolic region as indicated in Fig. \ref{Fig:accond_pristine}. This can be traced to the resonant-like feature in the optical absorption $\text{Re}(\sigma_{yy})$ at $\omega_{res} = 2.6$ eV. The spectral shape of $\text{Re}(\sigma_{yy})$ can be described by a Fano resonance curve \begin{equation}
    \sigma_{F,yy} \sim \frac{(q_F \Gamma_{res}/2 + \omega - \omega_{res})^2}{(\Gamma_{res}/2)^2 + (\omega - \omega_{res})^2}
\end{equation}
and setting $\sigma_{F,xx} \sim \omega^n$. A fit of these curves to the region around the second peak in the optical conductivity of bilayer BP is shown by the dashed lines in Fig. \ref{Fig:accond_pristine}. Its Kramers-Kronig pair, which correponds to $\text{Im}(\sigma_{yy})$, reveals a sign change after $\omega_{res}$. Hence, we can attribute the origin of hyperbolicity to the strong and anisotropic resonant-like interband transitions.

\begin{figure}[t]
\includegraphics[width=\linewidth]{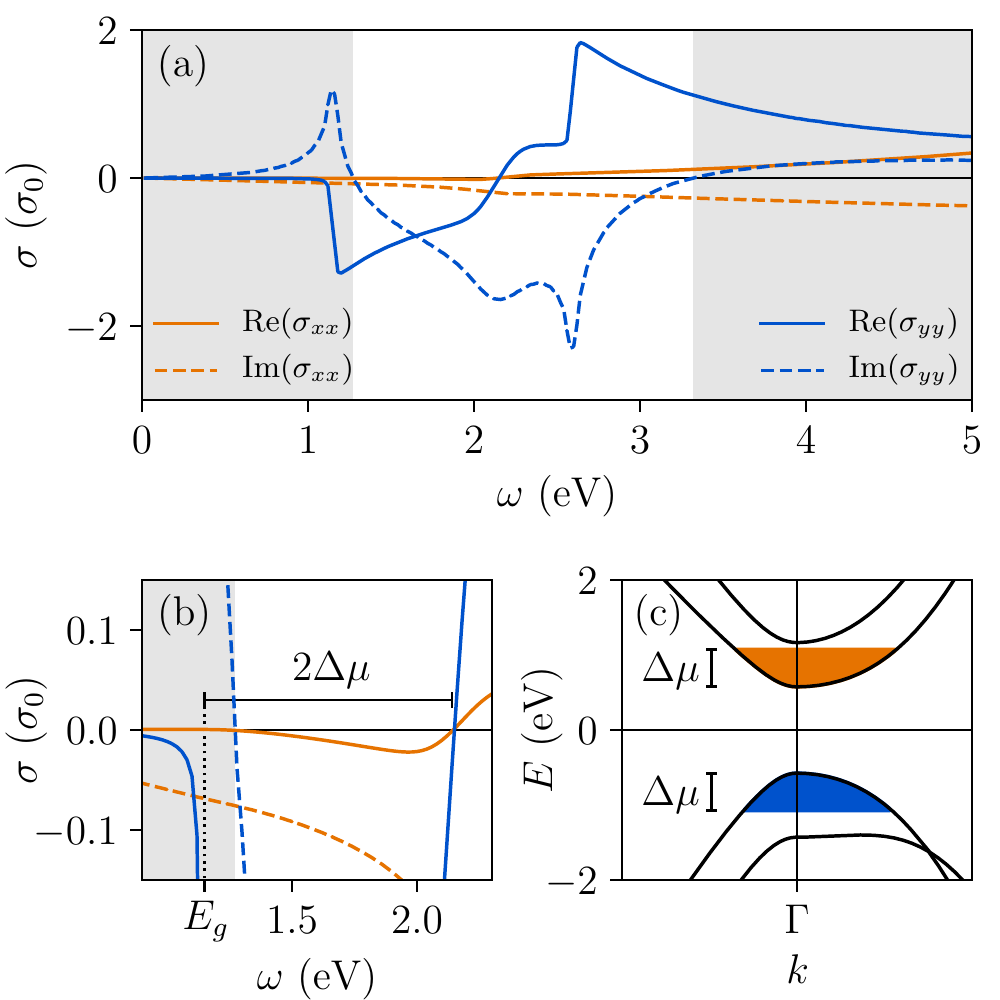}
\caption{(a) The optical conductivity of bilayer black phosphorus with photo-doping $\Delta \mu = 0.5$ eV. (b) A close-up of the region where $\text{Re}(\sigma_{yy}) < 0$, showing a new hyperbolic region (shaded) for $\omega < 1.27$ eV. (c) The corresponding band structure around the  $\Gamma$-point, with the population-inverted pockets shown in blue and orange.
}
\label{Fig:accond_gain}
\end{figure}

If we introduce population inversion (Eq. \ref{Eq:gain}, Fig. \ref{Fig:accond_gain}), however, the situation becomes qualitatively different: a new hyperbolic region appears in the infrared range. Here, we assume that the quasi-Fermi levels are such that
$\mu_h = −\mu_e$, and that the electron and hole baths can be described by a common temperature (see Fig. \ref{Fig:accond_gain}c). Optical pumping \cite{ni2016ultrafast,lui2010ultrafast}, where electrons and holes are generated in pairs, of a charge neutral system with particle-hole symmetry would fit such scenario. The optical gain causes $\text{Re}(\sigma_{yy})$ to become negative. The spectral window where $\text{Re}(\sigma_{yy})<0$ roughly coincides with $E_g<\omega<E_g+2\Delta\mu$, where there optical transitions between the population inverted electron and hole bands are allowed. In the region up to $\omega = 1.27$ eV, we find that $\text{Im}(\sigma_{yy}) > 0$. In the zigzag direction, the real part of the conductivity also flips sign between $E_g<\omega<E_g+2\Delta\mu$. The imaginary part in this direction, on the other hand, remains negative throughout the entire frequency range, causing a new hyperbolic region in the infrared.

\begin{figure}[t]
\includegraphics[width=\linewidth]{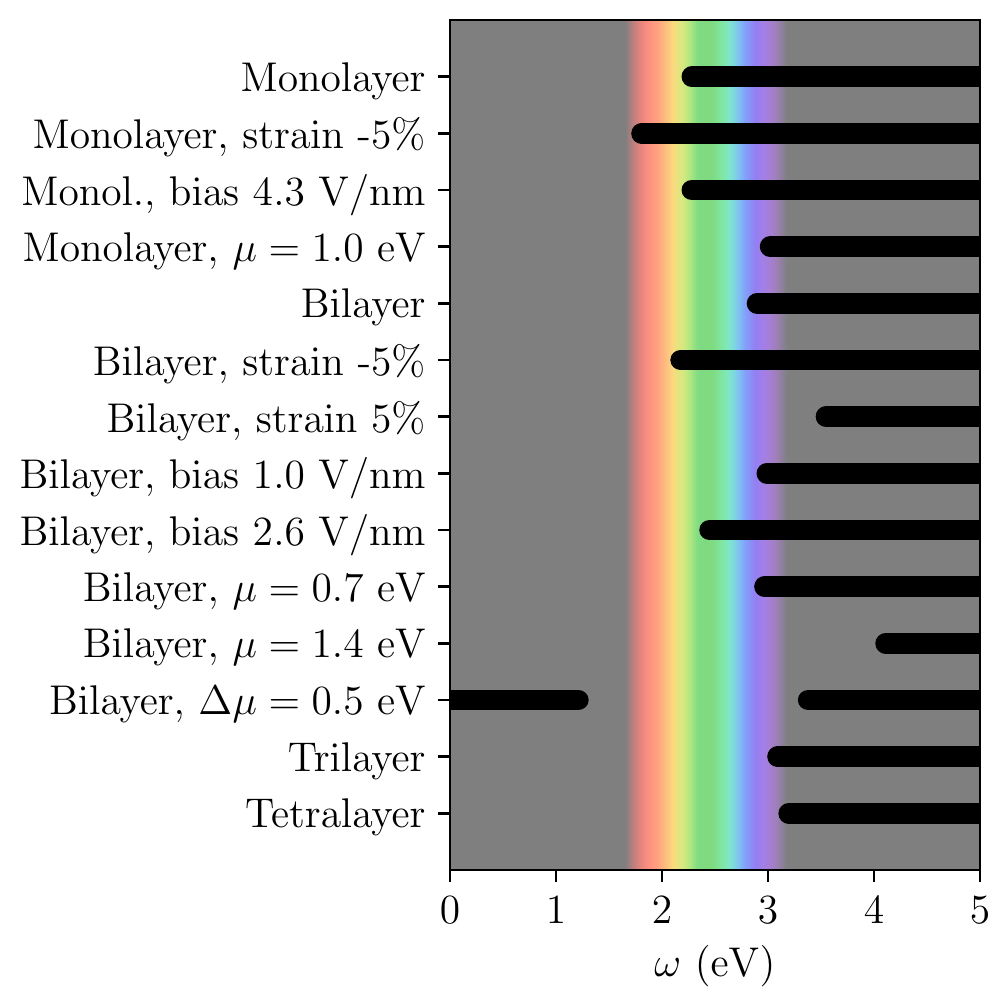}
\caption{The hyperbolic region (indicated in black lines) for different tuning parameters. The visual spectrum is indicated in color.}
\label{Fig:hyperbolic_regions}
\end{figure}

Since the origin of the hyperbolicity is related to the strong resonant-like anisotropic interband absorption between the largest conduction and valence subband indices, one expects that the spectral range of hyperbolicity can be tuned with bandstructure engineering. Indeed, the onset of the hyperbolic region $\omega_h$ can be tuned with the number of layers, strain, bias and doping (Fig. \ref{Fig:hyperbolic_regions}). $\omega_h$ goes down for compressive strain, because the peaks in the optical spectrum move to lower frequencies \cite{quereda2016strong}. For increasing bias in the bilayer case, the hyperbolic onset frequency $\omega_h$ first goes up, because the bands corresponding to the excitation causing the second peak move away from one another. Then it goes down as the band gap closes and a new peak appears in between the two existing peaks, because breaking the mirror symmetry in the $z$-direction allows for new hybrid transitions \cite{zhang2017infrared}. For more details about how the optical conductivity is affected by these parameters, we refer to the supplemental material. Moreover, $\omega_h$ goes up for an increasing number of layers, because extra layers add peaks to the optical conductivity in the armchair direction, and the hyperbolic region appears after the last peak. Finally, for increased doping, $\omega_h$ moves further up, as the first peak becomes less prominent due to Pauli blocking.

\emph{Hyperbolic Plasmons.}
Finally, let us consider characteristics of plasmons that can be supported by an hyperbolic material. We assume that the plasmon propagates at an angle $\chi$ with respect to the $x$-axis, where the plasmon wavector takes the form $\mathbf{q} = q_x \mathbf{e}_x + q_y \mathbf{e}_y + i\mathbf{e}_z \gamma$, where $q_x = q_{\parallel} \cos\chi$, $q_y = q_{\parallel} \sin\chi$, $q_{\parallel} = \sqrt{q_x^2 + q_y^2}$. The dispersion relation for the hyperbolic plasmon takes form
\begin{align}  \label{Eq:solution_dispersion}
   q_{\parallel}^2 = \gamma^2 + k_0^2, 
\end{align}
where
\begin{align} \label{Eq:gamma}
\gamma & =  \frac{i k_0}{2\sigma_{\mathbf{q}\mathbf{q}}} \left[ \left( \frac{2}{\eta_0} +\frac{\eta_0}{2} \sigma_{xx} \sigma_{yy} \right)\right. \notag \\
&+ \left. \sqrt{ \left(\frac{2}{\eta_0} + \frac{\eta_0}{2} \sigma_{xx} \sigma_{yy} \right)^2  - 4  \sigma_{\mathbf{q}\mathbf{q}}\sigma_{\perp\perp}} \right] ,
\end{align}
$k_0^2 = \omega^2 \mu_0 \varepsilon_0$, and
\begin{align}
& \sigma_{\mathbf{q}\mathbf{q}}(\chi) = \sigma_{xx} \cos^2 \chi +  \sigma_{yy} \sin^2\chi, \label{Eq:tensor_angle1} \\
& \sigma_{\perp \perp}(\chi) = \sigma_{xx} \sin^2 \chi  +  \sigma_{yy} \cos^2 \chi. \label{Eq:tensor_angle4}
\end{align}

\begin{figure}[t]
\includegraphics[width=\linewidth]{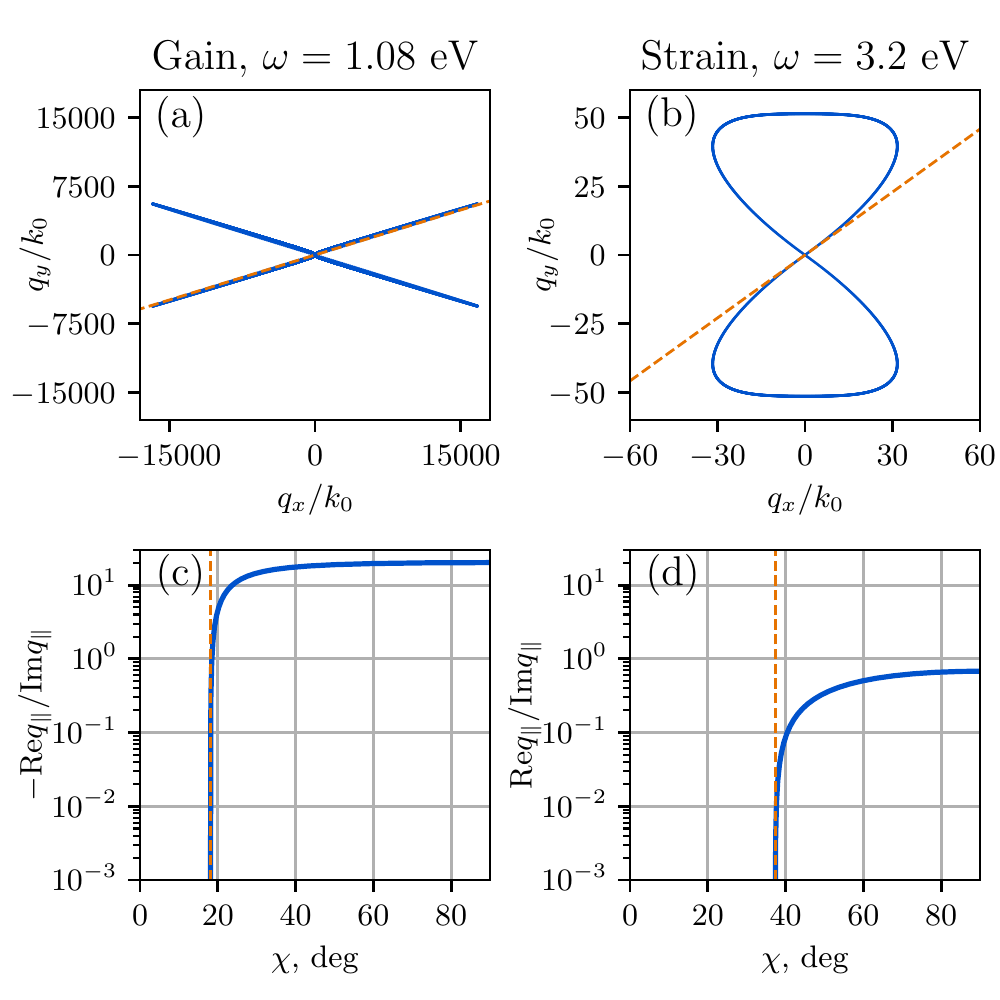}
\caption{Iso-frequency contours and figures of merit for hyperbolic materials with gain ($\Delta \mu = 0.5$ eV) and strain ($\epsilon_{yy} = -5\%$), calculated using Eq. \eqref{Eq:solution_dispersion}. The hyperbola asymptotes (orange dashed lines) are defined using Eq. \eqref{Eq:as}.}
\label{Fig:k_surfaces}
\end{figure}

The iso-frequency contours ($\omega(q_x, q_y) = \mathrm{constant}$), calculated using Eq. \eqref{Eq:solution_dispersion}, are presented in Fig. \ref{Fig:k_surfaces}. We consider the cases of BP with gain ($\sigma_{xx} = (0.07 -i 6.4) \times 10^{-2} \sigma_0$, $\sigma_{yy} = (-0.03+i0.60) \sigma_0$) and BP under strain ($\sigma_{xx} = (0.20 -i0.31) \sigma_0$, $\sigma_{yy} = (0.20 + i 0.54) \sigma_0$). It can be seen from Figs. \ref{Fig:k_surfaces}a,b, that only in the case of BP with gain the iso-frequency contour resembles a hyperbola with the asymptotes defined as 
\begin{align} \label{Eq:as}
 \tan \chi_0 = \sqrt{\left|\frac{\mathrm{Im} (\sigma_{xx})}{\mathrm{Im}(\sigma_{yy})}\right|}.
\end{align}
For BP under strain, the iso-frequency contour resembles a figure eight shape, even though the hyperbolicity condition \eqref{Eq:hyperbolic_condition} is met. 

This behavior stems from the fact that the hyperbolic shape of the iso-frequency contour is related to the poles of the denominator in Eq. \eqref{Eq:gamma}, defined by zeros of $\sigma_{\mathbf{q}\mathbf{q}}$ (i.e.\ when $\sigma_{\mathbf{qq}} \to 0$, then $|\gamma| \to \infty$ and $|q_{\perp}| \to \infty$). In particular, it is straightforward to demonstrate that in the case of a purely imaginary conducitivity tensor (i.e.\ no losses or gain) the condition $\sigma_{\mathbf{q}\mathbf{q}} = 0$ leads to Eq. \eqref{Eq:as} for hyperbola asymptotes. 

If the components of the conductivity tensor are both lossy ($\mathrm{Re}(\sigma_{xx,yy}) > 0$) or both have gain ($\mathrm{Re}(\sigma_{xx,yy}) < 0$), the module of the conductivity, $|\sigma_{\mathbf{qq}}|$, is never zero. In fact, for the hyperbola asymptote angle $\chi_0$, $\left|\sigma_{\mathbf{qq}}(\chi_0)\right| = \left|\mathrm{Re}(\sigma_{xx})\right| \cos^2 \chi_0 + \left|\mathrm{Re}(\sigma_{yy})\right| \sin^2\chi_0$. Thus, $\left|\sigma_{\mathbf{qq}}(\chi_0)\right|$ increases with the increase of $\left|\mathrm{Re}(\sigma_{xx, yy})\right|$ both for the lossy material and the material with gain. This leads to the decrease of $q_{\parallel}(\chi_0)$ and, eventually, destroys hyperbolicity when either losses or gain are too high. For example, this is the case in BP with strain presented in Figs. \ref{Fig:k_surfaces}b,d. The high losses of the material in the hyperbolic regime lead to the hyperbola folding into a figure eight shape iso-frequency contour. Moreover, the plasmons itself are very lossy in this case as is quantified by ratio $\mathrm{Re} q_{\parallel}/\mathrm{Im} q_{\parallel}$ in Fig. \ref{Fig:k_surfaces}d. 

The case where one of the components of the conductivity tensor is lossy ($\mathrm{Re}(\sigma_{xx}) > 0$, while the other has gain ($\mathrm{Re}(\sigma_{yy}) < 0$), requires separate consideration. In this case, $\sigma_{\mathbf{qq}}(\chi_0) = \left|\mathrm{Re}(\sigma_{xx})\right| \cos^2 \chi_0 - \left|\mathrm{Re}(\sigma_{yy})\right| \sin^2\chi_0$. When both the losses and the gain are small, then $\left|\sigma_{\mathbf{qq}}(\chi_0)\right|$ is small as well which allows for the iso-frequency contour to preserve the hyperbolic shape, as is the case for BP with gain presented in Fig. \ref{Fig:k_surfaces}a,c. This is, however, a rather trivial case which can be observed in pure lossy materials when the losses are small \cite{Nemilentsau_PRL_2016}. A non-trivial property of a material with gain is that $\sigma_{\mathbf{qq}} = 0$, when $\tan\chi_0 = \sqrt{\left|\mathrm{Re}(\sigma_{xx})\right| / \left|\mathrm{Re}(\sigma_{yy})\right|} $. That condition, together with Eq. \eqref{Eq:as}, indicates that the iso-frequency contour preserves its hyperbolic shape for arbitrary large losses and gain, as long as the following holds true,
\begin{equation} \label{Eq:hyperbola_gain}
    \sigma_{xx} = - \sigma_{yy} \tan \chi_0. 
\end{equation}
This criterion can be verified through its iso-frequencies contours, which we defer to the SI. Hyperbolic metasurfaces also offer the possibility of spontaneous emission rate enhancement through the Purcell effect \cite{PhysRev.69.681,Gjerding2017}. We defer these results to the SI. 

\emph{Conclusion.} In conclusion, we showed in this work how BP can be made hyperbolic across a broad spectral range, and studied the influence of optical gain on the hyperbolic plasmons. The ease of electrical tuning of the plasmons would open up new opportunities for flatland nanophotonics .

\emph{Acknowledgements.}
RR acknowledges financial support from the Spanish MINECO through Ram\'on y Cajal program,  Grants No.\ RYC-2016-20663 and Grant No.\ FIS2014-58445-JIN. MIK acknowledges financial support from the European Research Council Advanced Grant program (Contract No.\ 338957). SY acknowledges financial support from National Key R$\And$D Program
of China (Grant No. 2018FYA0305800) and National Science
Foundation of China (Grant No. 11774269).

\appendix{}

\bibliography{bibliography}

\end{document}


\title{Supplemental Material: Tuning 2D hyperbolic plasmons in black phosphorus}

\author{Edo van Veen}
\affiliation{School of Physics and Technology, Wuhan University, Wuhan 430072, China}
\affiliation{Institute for Molecules and Materials, Radboud University, Heyendaalseweg 135, 6525AJ Nijmegen, The Netherlands}
\author{Andrei Nemilentsau}
\affiliation{Department of Electrical and Computer Engineering, University of Minnesota, Minneapolis, Minnesota 55455, United States}
\author{Anshuman Kumar}
\affiliation{Physics Department, Indian Institute of Technology Bombay, Mumbai 400076, India}
\author{Rafael Rold\'an}
\affiliation{Materials Science Factory. Instituto de Ciencia de Materiales de Madrid (ICMM), Consejo Superior de Investigaciones Cient\'{i}ficas (CSIC), Cantoblanco E28049
Madrid, Spain}
\author{Mikhail I. Katsnelson}
\affiliation{Institute for Molecules and Materials, Radboud University, Heyendaalseweg 135, 6525AJ Nijmegen, The Netherlands}
\author{Tony Low}
\email{tlow@umn.edu}
\affiliation{Department of Electrical and Computer Engineering, University of Minnesota, Minneapolis, Minnesota 55455, United States}
\author{Shengjun Yuan}
\email{s.yuan@whu.edu.cn}
\affiliation{School of Physics and Technology, Wuhan University, Wuhan 430072, China}
\affiliation{Institute for Molecules and Materials, Radboud University, Heyendaalseweg 135, 6525AJ Nijmegen, The Netherlands}

\date{\today}

\maketitle

\begin{figure}[ht]
\includegraphics[width=0.49\linewidth]{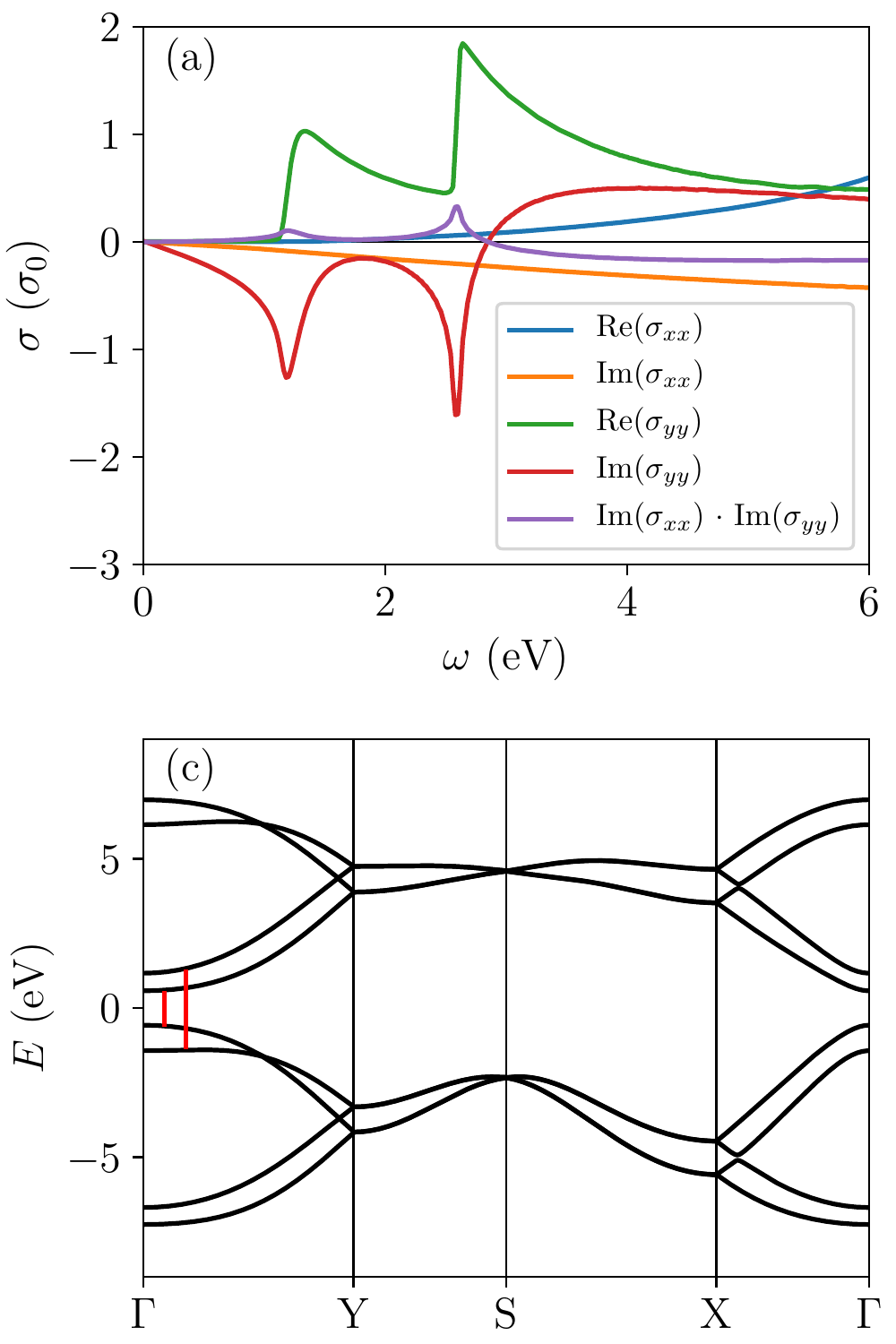}
\includegraphics[width=0.49\linewidth]{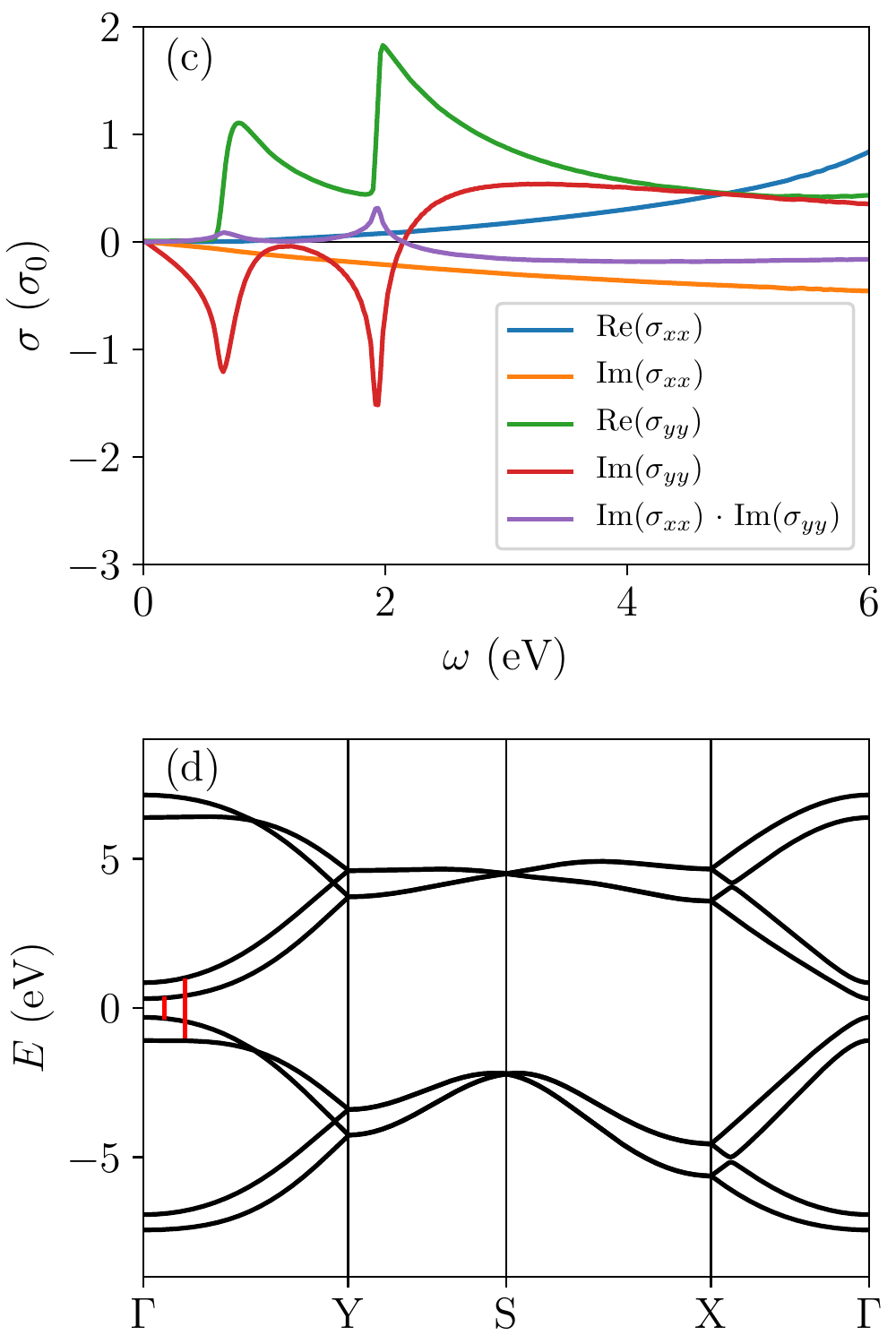}
\caption{The optical conductivity of bilayer black phosphorus (a) without strain; (b) with strain $\epsilon_{yy} = -5\%$. The corresponding band structures are given in (c) and (d), respectively, with optical excitations indicated in red.}
\label{Fig:SIstrain}
\end{figure}

\begin{figure}[ht]
\includegraphics[width=0.49\linewidth]{figures/SIfig0.pdf}
\includegraphics[width=0.49\linewidth]{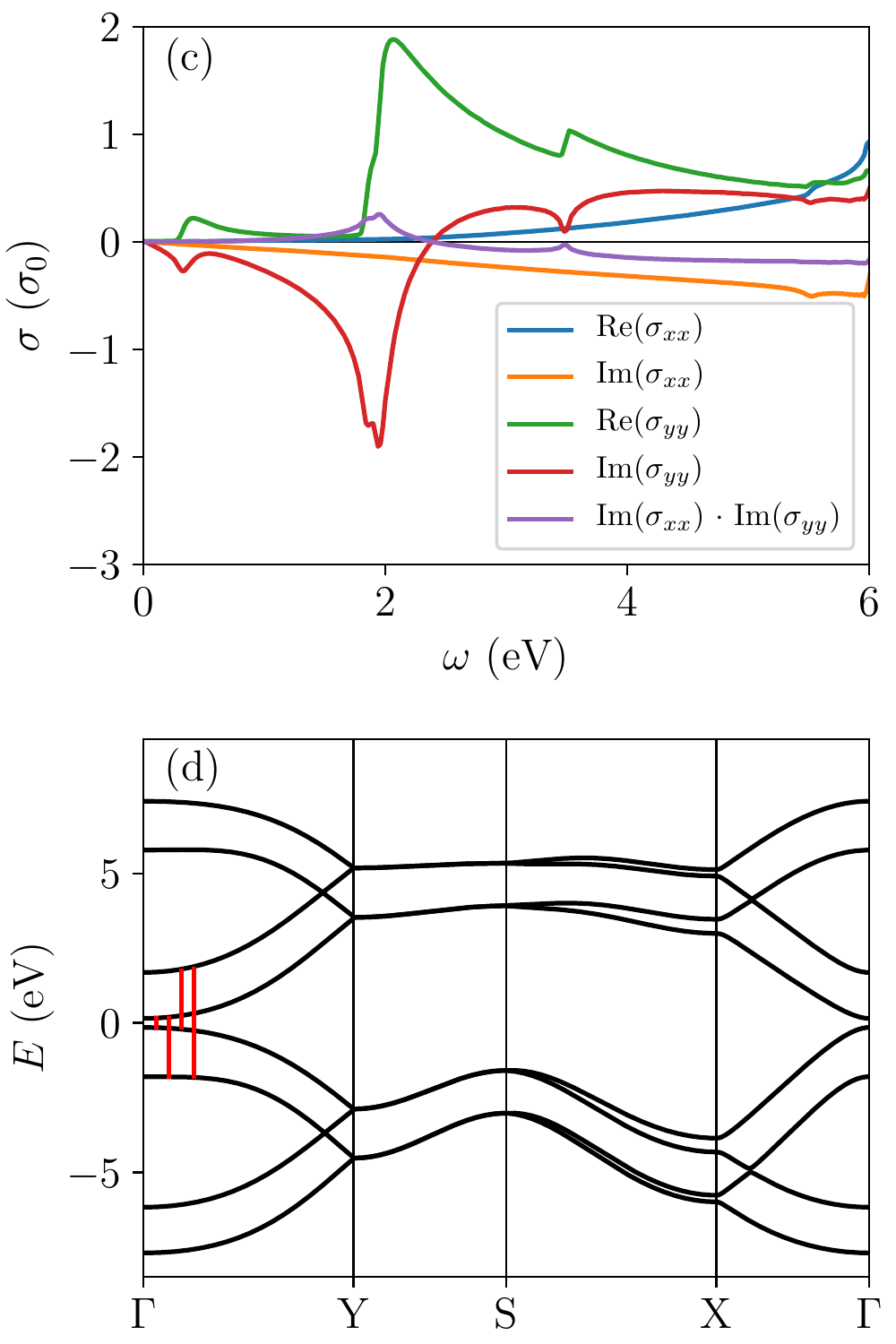}
\caption{The optical conductivity of bilayer black phosphorus (a) without bias; (b) with a bias of 2.6 V/nm. The corresponding band structures are given in (c) and (d), respectively, with optical excitations indicated in red.}
\label{Fig:SIbias}
\end{figure}

\emph{Tuning the optical conductivity: }Under compressive strain (Fig.\ \ref{Fig:SIstrain}) the band gap of bilayer black phosphorus becomes smaller. As a result, the optical peaks shift to lower frequencies. The introduction of bias (Fig.\ \ref{Fig:SIbias}) breaks the mirror symmetry in the $z$-direction, which allows for new hybrid transitions, as indicated in the band structure (Fig.\ \ref{Fig:SIbias}d). Moreover, the bands closest to the Fermi energy get pulled closer together, but the next pair of bands get pushed away from one another. This causes the hyperbolic region to go up at first, and then move to lower frequencies as the new peak in the middle gains amplitude.

\emph{Purcell factors: }When an excited emitter is placed near a nanostructure supporting a photonic mode, the rate at which the former gets rid of its energy is modified compared to the case when the emitter was in free space. This phenomenon is referred to as Purcell enhancement or spontaneous emission enhancement. Since bilayer black phosphorus supports hyperbolic plasmon modes, we present the possibility of Purcell enhancement in this system.
\begin{figure}[ht]
\includegraphics[width=\linewidth]{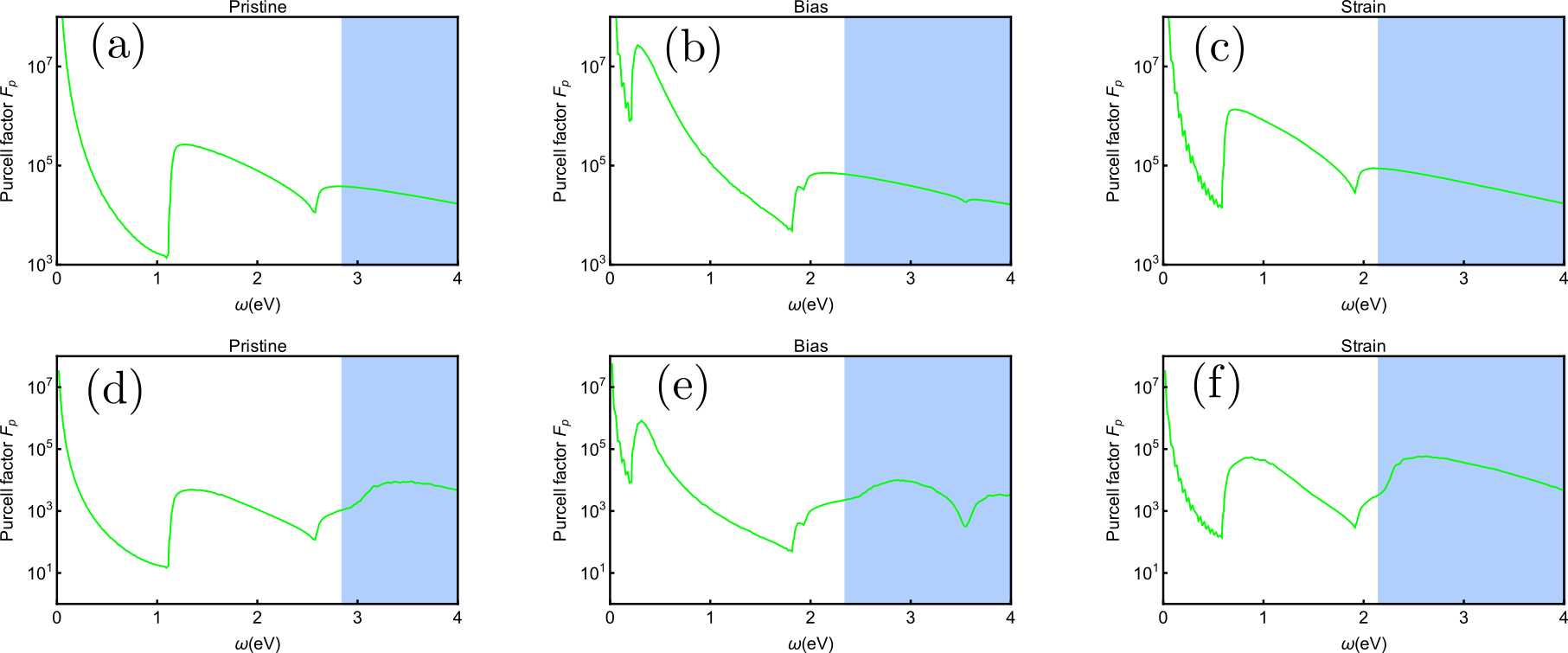}
\caption{\textit{Upper panel (a), (b), (c) :} Purcell enhancement for a $z-$polarized emitter placed 1 nm above the 2D sheet without gain. Elliptical to hyperbolic transition is not observed in the Purcell spectra due to the high interband losses. \textit{Lower panel (d), (e), (f):} In the low loss regime (loss reduced by 100), one can identify this transition.}
\label{Fig:Purcell}
\end{figure}
In order to study spontaneous emission rate (SER) engineering in an anisotropic 2D material, we employ the formalism of Ref.~\cite{Gomez-Diaz:15} where:
\begin{equation}
\textrm{SER} = \frac{P}{P_0} = 1 + \frac{6\pi}{k_0}\hat{\mathbf{n}}\cdot\Im\{\rttensor{G}_s(\mathbf{r_0},\mathbf{r_0};\omega)\}\cdot\hat{\mathbf{n}}
\end{equation}
If we consider only the vertically polarized dipole emitter (i.e. polarized out of the plane of the BP sheet) placed a distance $d$ above the sheet, the scattered Green's tensor simplifies to:
\begin{align}
&\hat{\mathbf{n}}\cdot\{\rttensor{G}_s(\mathbf{r_0},\mathbf{r_0};\omega)\}\cdot\hat{\mathbf{n}} \nonumber \\
&= \frac{\imath}{8\pi^2}\int\int\Gamma_{\textrm{pp}}\hat{\mathbf{n}}\cdot\rttensor{M}_{\textrm{pp}}\cdot\hat{\mathbf{n}}e^{2\imath k_z d}dk_xdk_y
\label{eq:Greens}
\end{align}
Here the two terms in the integrand are given by:
\begin{equation}
\hat{\mathbf{n}}\cdot\rttensor{M}_{\textrm{pp}}\cdot\hat{\mathbf{n}} = \frac{k_{\rho}^2}{k_z k_0^2}
\end{equation}
and 
\begin{equation}
\Gamma_{\textrm{pp}} = \frac{-Z_0\sigma_{xx}' (2Z^s+Z_0\sigma_{yy}')+Z_0^2\sigma_{xy}'\sigma_{yx}'}{(2Z^s+Z_0\sigma_{yy}')(2Z^p+Z_0\sigma_{xx}')-Z_0^2\sigma_{xy}'\sigma_{yx}'}
\end{equation}
with $Z_0=\sqrt{\mu_0/\epsilon_0}$, $Z^s = k_z/k_0$, $Z^p = k_0/k_z$ and $\rttensor{\sigma}'$ denotes the rotated surface conductivity tensor. If the material has no intrinsic off-diagonal contribution to the surface conductivity, then the components of the rotated surface conductivity tensor simplify to:

\begin{align}
\sigma_{xx}' &= \frac{k_x^2\sigma_{xx} + k_y^2\sigma_{yy}}{k_{\rho}^2} \nonumber \\
\sigma_{yy}' &= \frac{k_x^2\sigma_{yy} + k_y^2\sigma_{xx}}{k_{\rho}^2} \nonumber \\
\sigma_{xy}' &= \sigma_{yx}' = \frac{k_x k_y (\sigma_{yy} - \sigma_{xx})}{k_{\rho}^2} \nonumber 
\end{align}
The result of this calculation is presented in Fig.~\ref{Fig:Purcell}, where we assume a $z-$polarized emitter located at a distance of 1nm from free standing bilayer black phosphorus. Due to the large loss in the optical conductivity in the hyperbolic regions, a clear transition from the elliptical to hyperbolic regime\cite{Gjerding2017} is not observed, as shown in Fig.~\ref{Fig:Purcell} (a), (b), (c). The fact that this is due to the large losses is confirmed by carrying out similar calculations where the real part of the optical conductivity is reduced by a multiplicative factor of 100. In this case one can clearly observe the elliptical to hyperbolic transition as shown in Fig.~\ref{Fig:Purcell} (d), (e), (f). Moreover, the transition point of the Purcell factor is clearly shown to be tunable as a function of bias and strain.

\begin{figure}[t]
\includegraphics[width=\linewidth]{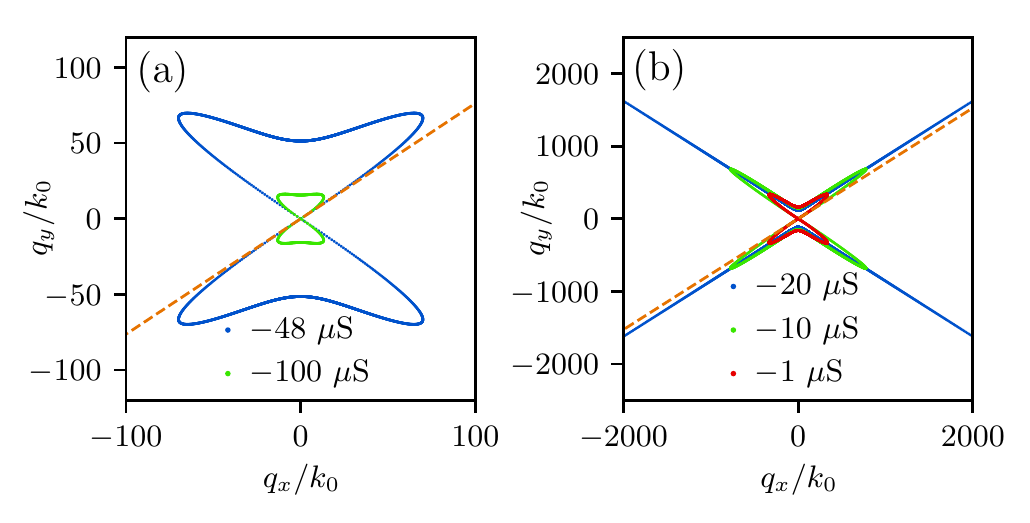}
\caption{Iso-frequency contours for hyperbolic materials with different values of gain. $\sigma_{xx} = 12 -i19$ $\mu$S, $\sigma_{yy} = \mathrm{Re}(\sigma_{yy}) + i 33$ $\mu$S). $\mathrm{Re}(\sigma_{yy})$ is indicated in the legend on the panels (a), (b). }
\label{Fig:k_surfaces_gain}
\end{figure}

\emph{Hyperbolic plasmons with gain: } We demonstrate that equation 
\begin{equation} \label{Eq:hyp_gain}
    \sigma_{xx} = - \sigma_{yy} \tan \chi_0. 
\end{equation}
is indeed beneficial for the case of the hyperbolic plasmons when one of the components of the conductivity tensor has gain, while the other one is lossy,
by considering iso-frequecy contours for the materials with different values of gain. To illustrate this, we choose BP under strain, and assume that a gain was added to $y$-component of the BP conductivity, with the gain values ($\mathrm{Re}(\sigma_{yy})$) indicated in the legend of Fig. \ref{Fig:k_surfaces_gain}. Due to high losses, the iso-frequency contour of BP under strain does not resemble a hyperbola. Countering the high losses with high gain is not beneficial for restoring the hyperbolic plasmons (see Fig. \ref{Fig:k_surfaces_gain}a) as this leads to a high magnitude of $\sigma_{\mathbf{qq}}$. However, we can recover the hyperbolic mode by matching $\sigma_{xx}$ to $\sigma_{yy}$ using Eq. \eqref{Eq:hyp_gain}, as can be see from Fig. \ref{Fig:k_surfaces_gain}b ($\mathrm{Re}(\sigma_{yy}) = - 20 \mu$S). A further decrease of gain breaks the resonance condition \eqref{Eq:hyp_gain} and leads to distortion of the hyperbola ( $-10$ $\mu$S and $-1$ $\mu$S in the Fig. \ref{Fig:k_surfaces_gain}b).

\bibliography{bibliography}